\documentclass[11pt]{article}
\usepackage{moriond,epsfig}

\def\ra{\rightarrow}

\def\be{\begin{equation}}
\def\ee{\end{equation}}
\def\bea{\begin{eqnarray}}
\def\eea{\end{eqnarray}}

\def \Et {{\rm E}_{\rm T}}
\def \Pt {{\rm P}_{\rm T}}
\newcommand{\MET}{\mbox{$\protect \raisebox{.3ex}{$\not$}\Et$}}
\begin{document}
\vspace*{4cm}
\title{EARLY LHC PHYSICS STUDIES: WHAT CAN BE OBTAINED BEFORE DISCOVERIES?}

\author{ K. LASSILA-PERINI }

\address{Helsinki Institue of Physics,  P.O.Box 64 (Gustaf H\"allstr\"omin katu 2),\\
 FIN-00014 University of Helsinki, Finland}

\maketitle\abstracts{
The Large Hadron Collider will provide an unprecedented quantity of collision data
right from the start-up. The challenge for the LHC experiments is the quick use of these data
for the final commissioning of the detectors, including calibration, alignment, 
measuring of detector and trigger efficiencies. A new energy frontier will open up,
and measurement of basic Standard Model processes will build a solid basement for any
discovery studies.} 

\section{Introduction}
Apart from the well studied discovery potential for missing pieces in the Standard Model,
supersymmetric particles and new physics, the Large Hadron Collider (LHC) will provide
unprecedented opportunities to explore the frontier of high energy physics.
Excellent reports have been presented considering the start-up strategy for the
LHC experiments ATLAS and CMS~\cite{start-up-strategy}, this note aims to concentrate
on questions such as
\begin{enumerate}
\item what will the early LHC data consist of?
\item what will be the first physics publications of the LHC?
\item what is the work required before any results can be published?
\end{enumerate}
These questions will be addressed by a hypothetical and speculative list
of first LHC papers.

\section{The context}
The LHC is expected to provide collisions at nominal luminosity values of
2$\cdot$10$^{33}$ cm$^{-2}$s$^{-1}$ (``low luminosity'') and 
10$^{34}$ cm$^{-2}$s$^{-1}$ (``high luminosity''). During
the low luminosity running the experiments are expected to collect
10-20 fb$^{-1}$ of data per year, whereas the high luminosity running
will yield 100 fb$^{-1}$ of collected data yearly.
The nominal values will be reached after a period of pilot runs
during the accelerator and physics commissioning where
the machine and detector conditions will be studied and understood.
This note concentrates on these early periods of LHC running, and the
following naming convention is used:
\begin{itemize}
\item {\bf Pilot run}: machine development run interleaved with data-taking runs
\item {\bf First physics run}: run where the nominal machine parameters and the nominal low luminosity 
will be gradually reached, but the
integrated luminosity is limited by the time needed to master the LHC operation. 
\end{itemize}
With the present knowledge, the pilot run will start in 2007 and it can be estimated to take one month,
the first physics run will follow the year after. For the pilot run, the integrated luminosity recorded
per experiment is estimated to be of order of 10 pb$^{-1}$ if an average luminosity
of \mbox{10$^{31}$ cm$^{-2}$s$^{-1}$} is reached. The integrated luminosity of the first physics run
is envisaged to be of order of some fb$^{-1}$.

The discovery potential of the LHC detectors has been thoroughly studied and will not be discussed
in this note. It is useful to remember, however, that already with 1 fb$^{-1}$ of collected data
a wide range of the possible Standard Model (SM) Higgs boson mass reach can be covered and
that the experiments have a good sensitivity to supersymmetric particles, many of which can
be discovered or excluded with the early data.
In most discovery areas, nevertheless, a good understanding of the detector performance is required
and the background processes need to be well understood before claiming discovery or exclusion, so it
is most likely that such discovery papers are not among the earliest LHC publications, except
if Nature has reserved surprises just over the high energy corner.

\section{Start-up data}
Due to the significant increase of the cross-sections of the hard processes
as a function of collision energy,
the event rates at the LHC start-up will be larger than other colliders
even with a modest luminosity. To reach the before mentioned 10 pb$^{-1}$ of recorded data during the pilot run, 
a fairly low data taking efficiency of 20\% (luminosity recorded/luminosity delivered) 
is assumed at the start-up.
As a fictive example of the
possible start-up rates, 
a selection efficiency of 20\% (including
the geometrical acceptance, $p_t$ cuts, and detector efficiency)
is assumed for leptons from $W \rightarrow \ell\nu$ and $Z \rightarrow \ell\ell$
and 1.5\% for $t\bar{t} \rightarrow \ell\nu + X$. Such selection efficiencies
are much lower that foreseen in the two experiments and they are meant
to be pessimistic including anything that can go wrong at the start-up.
Possible event samples collected in one mont of a pilot run 
are shown in Table~\ref{tab:rates}. Even with these pessimistic assumptions,
the very first LHC operation will provide a large statistics of interesting
data at a completely new energy frontier. The challenge for the LHC
experiments is to make best profit of these data and complete the
mapping of the detector performance and make the very first
physics measurements.

\begin{table}[h]
\caption{Expected data samples from the pilot run.\label{tab:rates}}
\vspace{0.4cm}
\begin{center}
\begin{tabular}{|l|l|l|l|}
\hline

{\bf Process} & 
{\bf $\sigma \times$ BR} &
{\bf $\varepsilon$ (estimate)} &
{\bf Events selected in 10 pb$^{-1}$} \\ \hline
$W \rightarrow \ell\nu$ & 20 nb & $\sim$20\% & $\sim$ 40000\\ 
$Z \rightarrow \ell\ell$ & 2 nb & $\sim$20\% & $\sim$ 4000\\
$t\bar{t} \rightarrow \ell\nu + X$ & 370 pb & $\sim$1.5\% & $<$ 100 \\
Jet $E_t > 25$ GeV & 3 mb & 100\% &  $\sim$3$\cdot$10$^{10}\times$ prescale factor \\
Jet $E_t > 140$ GeV & 440 nb & 100\% &  $\sim$4.4$\cdot$10$^{6}$ \\
Minimum bias & 100 mb &  & 10$^{12}\times$ prescale factor \\ \hline
\end{tabular}
\end{center}
\end{table}

\section{The experiments at the start-up}

At the moment of writing, the construction of the LHC experiment is proceeding
at full speed. The ATLAS experiment is being built in its final position
in the experimental pit and the lowering of the large components of the CMS experiment 
from the construction hall at the surface to
the experimental cavern
is being prepared. At the start-up, the detectors will
be fairly complete and ready enough for early physics.
Some items, however, will not be present during the pilot run and they
will be installed during the first long
accelerator shut-down between the pilot run and the first physics run.

The final alignment and calibration will only take place with the real
data and, therefore, at the very beginning, there are somewhat large
uncertainties on the detector measurements, due to non optimal
uniformity, $e$/$\gamma$/jet scale and track alignment.

\section{The first publication: ``
Charged particle multiplicity in pp collision at $\sqrt{s}$ = 12 TeV
and $\sqrt{s}$ = 14 TeV''}
\begin{quotation}
\noindent We report on a measurement of the mean charged particle multiplicity in minimum bias events, produced
at the LHC in pp collisions with $\sqrt{s}$ = 12 TeV and $\sqrt{s}$ = 14 TeV, and 
recorded in the CMS experiment at CERN.
The events have been selected by a minimum bias trigger, the charged tracks reconstructed in the
silicon tracker and in the muon chambers. The track density is compared to the results of Monte Carlo
programs and it is observed that none of the programs describe all features of the data.\footnote{All ``abstracts'' 
quotated in this note are purely fictive, speculative and not
necessarily based on solid studies, and the phrasing may have been strongly influenced by existing publications.}
\end{quotation}

\subsection{Why?}
The most common collisions --- the so called minimum bias events --- have large uncertainties, and
being soft processes, it is hard to predict their structure at the high LHC collision energies
Their measurement is vital for understanding the detector backgrounds, energy scales and
detector occupancy. Only when the minimum bias events are well known and understood, can the
reconstruction algorithm details be fixed.

\subsection{How to get there?}

Counting the charged tracks is a fairly simple measurement and doable already with an only
partly understood detector behaviour. Selecting the events can be done with a random trigger
firing on non-empty bunch crossing which can be identified from a scintillator or electromagnetic
calorimeter (ECAL) signal. Obviously, track reconstruction needs to be working, and to study
the very first collisions, it needs to work without the possibly missing detector element,
such as the inner pixel layers in the case of CMS.

Much of the work has already started before the first collisions. The experiments use
cosmic muons to verify the functioning of the track reconstruction already before the
final installation. Further on, in the operational position, the cosmic muons will be used to
align and calibrate the detector in the barrel region.
With the first LHC beam circulating, the beam halo muons --- machine induced secondary particles
crossing the detector horizontally --- can be used for alignment and calibration in the endcap
region. Still before the first collisions there will be beam gas interactions which produce 
collision like event structures with low $p_t$ tracks if interactions happen in the active detector volume.
All such data collected  before the start-up and during the single beam operation will be useful for gaining experience
with data taking, studying dead channels, debugging readout among others.

\begin{figure}[h]
\begin{center}
\psfig{figure=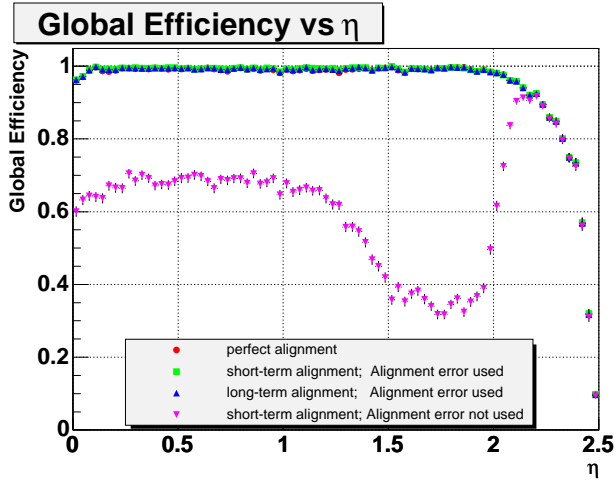,height=2.5in}
\end{center}
\caption{The effect of non perfect alignment on the track reconstruction efficiency
\label{fig:align}}
\end{figure}

The question to be asked for such an early measuremnt is how it is affected by the not yet
optimal detector performance. As the quantity to be measured is the number of
charged tracks rather than the track $p_t$, this study is fairly insensitive to
the alignment errors. As an example, Figure~\ref{fig:align} shows that  inefficiency on the global
track reconstruction induced by the rough alignment is negligible~\cite{align}. 

\section{Then follows: ``Measurement of inclusive jet cross section in pp collisions at \mbox{$\sqrt{s}$ = 12 TeV} and
 \mbox{$\sqrt{s}$ = 14 TeV}''?}

\begin{quotation}
\noindent We present results from the measurement of the inclusive jet cross section for jet transverse energies from 
100 to 1500 GeV in the pseudorapidity range 0.1 $<| \eta |< $1.4. The results are based on 18 pb$^{-1}$ of data 
collected by the ATLAS Collaboration at the Large Hadron Collider at CERN . The data are consistent with previously 
published results. The data are also consistent with QCD predictions given the flexibility allowed from 
current knowledge of the proton parton distributions. 
\end{quotation}

\subsection{Why?}
The pilot run will provide a large statistics of jet and single particle events at highest collision energies
ever measured. The statistical uncertainties on many measurements are quickly smaller than
experimental and theoretical systematic uncertainties. Studying jet events will yield information
of perturbative QCD and it will provide the first occasion to test the accuracy of parton
distribution function (PDF) whose uncertainty will influence many of the possible LHC measurements.  

\subsection{How to get there?}
Quoting cross-sections requires knowledge of connected quantities:

\begin{equation}
\sigma = n_{\mbox{\tiny events}}/(\varepsilon_{\mbox{\tiny trigger}} \cdot \varepsilon_{\mbox{\tiny selection}} \cdot {\textstyle \int} \cal{L})
\label{eq:sigma}
\end{equation}
where $\int \cal{L}$ is the integrated luminosity, $\varepsilon_{\mbox{\tiny selection}}$
the offline event selection efficiency and  $\varepsilon_{\mbox{\tiny trigger}}$ the online trigger
efficiency. The number of events $n_{\mbox{\tiny events}}$ is measured.

To present inclusive jet cross sections as suggested in this publication, 
the luminosity measurement is therefore needed. The goal at the LHC is to have the uncertainty
in this measurement lower that the theoretical uncertainty in cross-sections at the LHC
start-up which is estimated to be approximately 5\%.
The luminosity measurement consists of two equally important components, the luminosity determination
(the total p-p cross-section) and the luminosity monitoring (the instantaneous luminosity).
The former can determined with a special measurement in the very forward region and extrapolating
the total inelastic cross-section from the measured elastic cross-section according to
the optical theorem. This can only be done with a special beam optics different from that of
the normal running. Alternatively, the cross-section can be normalized to the SM predictions
of the W and Z cross-sections which are computed with 1-2\% precision.
The latter, the instantaneous luminosity, can be moritored by counting simple structures
in the detector, such as empty events or empty regions in the very forward calorimeter
or counting tracks. It is obvious that the precision on the integrated luminosity
at the start-up cannot be optimal, hardly any of the elements having been studied
in detail right at the beginning. Therefore, the induced uncertainty on a total cross-section
measurement will be large, but it should not prevent interesting studies on differential
cross-sections as a function of jet energy and jet position.

While the offline event selection efficiency is thoroughly quantified in the preparatory
studies on any physics signature, the trigger efficiency often draws less attention
by physicists, at least in the preparational phase of the experiments as now.
It is, however, one of the key factor in determining the cross-section, as shown
in Equation \ref{eq:sigma}.
The trigger efficiency varies as a function of energy and position of the triggered
objects, and it will be monitored by recording a fraction of rejected events and studying
the causes of rejection, and by recording prescaled trigger streams, i.e. recording
one in N (prescale factor) events with low trigger thresholds, and studying the efficiency
with which the events pass the higher threshold. The pilot run will provide enough
statistics for studying the trigger efficiency.

The measurement of differential jet cross-section requires knowledge of the jet energy.
The energy determination consists of three components: hadronic calorimeter (HCAL) calibration, jet energy scale,
and jet energy corrections. The {\em HCAL calibration} assures that
all elements give the same response to an equal signal. It can be achieved by
a scan by a radioactive source and by test beam calibration in a single particle beam.
A correct {\em jet energy scale} means that the measured jet energy corresponds to the
energy of the originating parton. This scale can be set by using jet events where the
jet energy is balanced by high $p_t$ particle (such as $\gamma$ or Z) in the opposite
direction. The high $t\bar{t}$ cross-section at the LHC allows $t\bar{t} \rightarrow
bW bW \rightarrow b\ell \nu jj$ events to be used to set the jet scale as W and top quark
masses are known. Isolated charged tracks may also be used to connect the single particle
beam calibration to the measured jet energy. The {\em jet energy corrections} are needed
to have a jet energy response linear in energy and independent of jet coordinates.
The corrections can be defined for example from di-jet events where one jet is in
a well calibrated energy and coordinate region and the other jet is to be calibrated.
While the pilot run will provide an excellent statistics for many jet studies, it
is likely that achieving the complete jet energy calibration requires some time
and detailed studies as many elements used in jet calibration need a calibration
of their own. 

One of the main issues of the jet energy spectrum measurement is to compare it with the 
SM predictions. There are many systematic uncertainties involved, as an 
example the long list studied by the experiments at Tevatron: high $p_t$ and low
$p_t$ hadron response, energy scale stability, underlying event, calorimeter
resolution, fragmentation of partons to stable hadrons, $e$/$\gamma$ response~\cite{tevjet}.
Large amount of work is needed to address all these issues which will be important
at the LHC as well, and it may well be that before the publication on jets
some single particle spectrum measurements or measurement of Drell-Yan production
of dileptons, and in particular $Z \ra \mu^+ \mu^-$ may make their way through as publications.

\section{And next: ``Search for new high mass particles decaying to lepton pairs in pp collisions at 
 \mbox{$\sqrt{s}$ = 14 TeV}''}

\begin{quotation}
\noindent A search for new particles ($X$) that decay to electron or muon pairs has been 
performed
using approximately 800 pb$^{-1}$ of~\mbox{$pp$}~collision
data at $\sqrt{s}=14$ TeV collected by the CMS experiment 
at the LHC at CERN. 
Limits on $\sigma(pp\rightarrow X) \cdot BR(X\rightarrow \ell\ell)$ are presented
as a function of dilepton invariant mass~$m(\ell\ell) > 150~\rm{GeV}/c^2$,
for different spin hypotheses (0, 1, or 2).  
Lower mass bounds for $X$~from representative models beyond the Standard Model 
including heavy neutral gauge bosons are presented.
\end{quotation}

\subsection{Why?}
New resonance in the di-lepton spectrum may be seen --- or excluded --- very early, already
at the beginning of the first physics run. There are many beyond SM scenarios which can
produce a new heavy di-lepton resonance, and among the possible early-comers are 
new heavy vector bosons~\cite{heavygb} or Randall-Sundrum gravitons~\cite{RSG}. They form 
a spectacular peak in absence of SM backgrounds.

\subsection{How to get there?}
In the most favourable cases, observation of heavy di-lepton resonances does not pose
particular problems. The measurement of a broad peak with low background is not sensitive
to detector uncertainties, and is therefore feasible already with early data.
However, when it comes to exclusion, more care is needed to quantify the systematic uncertainties.

Due to the large event rate at the LHC, the statistical uncertainty in many measurements soon
becomes negligible in comparison with the systematic uncertainties. The systematic
uncertainties may come from theoretical predictions or from the event modelling, they
may be purely experimental and be induced from the accelerator conditions.

\section{And certainly: ``Top quark mass measurement in pp collisions at \mbox{$\sqrt{s}$ = 14 TeV}'' }

\begin{quotation}
\noindent Preliminary results on the measurement of the top quark mass in the ATLAS 
experiment at the LHC are presented. 
In the lepton plus jets channel, the ATLAS Collaboration measures 178.2$\pm$0.4(stat.)$\pm$7.2(syst.)  GeV/$c^2$, 
using a sample of $\sim$102 pb$^{-1}$ at $\sqrt{s}=14$$~$TeV. Events with an isolated electron or muon with 
p$_t >$ 20 GeV/c and four jets with E$_t >$ 40 GeV were selected. 
The mass is obtained from the invariant mass of the three highest E$_t$ jets. 
The uncorrected mass shows a slight dependence on the jet energies, which has been taken into account in
the energy scale corrections.
\end{quotation}

\subsection{Why?}
The LHC is a top factory: the $t\bar{t}$ cross-section at the LHC energies is 833 pb while it is
7 pb at the Tevatron. In consequence, early LHC runs will provide high enough statistics to
observe and measure top quark properties with simple analysis techniques. 

\subsection{How to get there?}
\begin{figure}[h]
\begin{center}
\psfig{figure=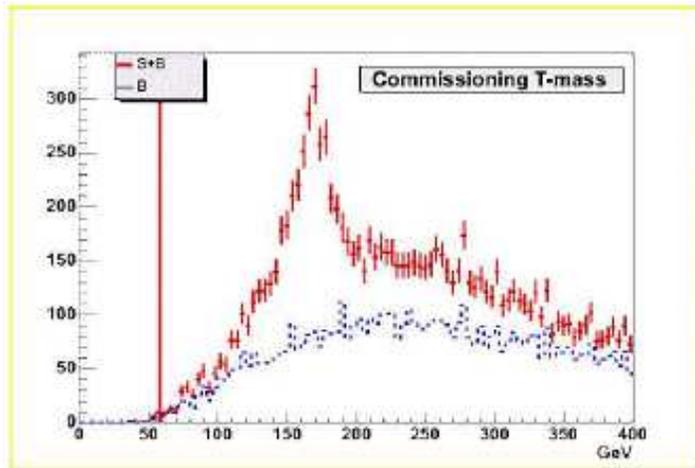,height=2.5in}
\end{center}
\caption{The reconstructed top mass, without $b$-tagging, for 150 pb$^{-1}$ of data. The $W$ + 4 jet
background is added to signal events and shown as a dashed line.
\label{fig:topmass}}
\end{figure}

The top quark mass at the LHC from can be first measured with a very simple and robust selection
of the $t\bar{t} \rightarrow bW bW \rightarrow b\ell \nu jj$ channel~\cite{atlastop}.
One isolated lepton ($e$ or $\mu$) is required and exactly 4 jets. No kinematic fits
nor $b$-tagging is required to start with, whereas these techniques are of ultimate
importance at the Tevatron due to the lower statistics. The mass peak can be obtained
by plotting the invariant mass of the highest $E_t$ jets as shown in Fig. \ref{fig:topmass}, 
and constraining two of the jets 
with the $W$ mass, the background can be further reduced.

Apart from being an important measurement on its own, the top mass measurement gives
important feedback on detector performance even at the early stage when the mass precision
is not yet competitive with earlier Tevatron studies. A wrong top mass
indicates error in the energy scale. The top sample will also be useful to
commission the $b$-tagging algorithms. 


\section{Conclusions}
Even with a limited performance at the start-up, the LHC will provide a large
quantity of data right from the beginning. Already the very first pilot run
can provide 10 pb$^{-1}$. These data will come in a short period of order of
one month and there will be very little time for fine-tuning. 
There are many unknowns and the machine operation will vary from single
beam to close to nominal conditions. The experiments will need to make
sure that as much useful data as possible will be recorded as the physics
commisioning of the detectors rely mainly on real collision data.

The first physics run is expected to provide some fb$^{-1}$ of data.
The data volumes are unprecedented and the collaborations will have
to cope with data access, fast reconstruction and analysis code
development cycle, and how to do all this worldwide, putting the
grid computing in serious use.

Despite of the difficulties, many interesting studies can be
made and will have to be made with the early data.
Examples of possible early measurements have been shown, and to conclude ---
although beyond the scope of this note --- yet another is suggested:

\section*{And perhaps: ``Evidence for squark and gluino production in pp collisions at \mbox{$\sqrt{s}$ = 14 TeV}'' }

\begin{quotation}
\noindent Experimental evidence for squark and gluino production in pp collisions \mbox{$\sqrt{s}$ = 14 TeV} with an integrated
luminosity of 97~pb$^{-1}$ at the Large
Hadron Collider at CERN is reported. The CMS experiment has collected 320 events of events with several
high $\Et$ jets and large missing $\Et$, and the measured effective mass, i.e. the scalar sum of the four highest $\Pt$
jets and the event $\MET$, is consistent with squark and gluino masses of the order of 650 GeV/$c^2$.
The probability that the measured yield is
consistent with the background is 0.26\%.
\end{quotation}

\section*{Acknowledgments}
This note relies on the work done by my CMS and ATLAS
colleagues. Many excellent presentations have been shown on this topic, and I'm grateful to
F. Gianotti, A. de Roeck, G. Rolandi and O. Buchmueller among others as I have learnt much on the subject
from their lectures and presentations. Furthermore, special thanks go to A. de Roeck, P. Janot, D. Denegri,
I. Tomalin and R. Tenchini for their comments and suggestions.

\end{document}